\documentclass[twocolumn,english,aps,prd,reprint,floatfix,notitlepage,footinbib,preprintnumbers,superscriptaddress]{revtex4-1}
\pdfoutput=1
\usepackage{lmodern}

\usepackage[T1]{fontenc}
\usepackage[latin9]{inputenc}
\usepackage{geometry}
\usepackage{subfigure,lmodern, amsmath,amssymb, graphicx, pifont, adjustbox, bm, xcolor}
\usepackage{amsfonts}
\usepackage{geometry}
\usepackage{comment}
\usepackage{mathtools}
\usepackage{float}
\usepackage{slashed}
\usepackage{ragged2e}
\usepackage{array}

\makeatletter\g@addto@macro\bfseries{\boldmath}\makeatother

\usepackage{stackengine}
\usepackage{esint}
\usepackage[unicode=true,pdfusetitle,
 bookmarks=true,bookmarksnumbered=false,bookmarksopen=false,
 breaklinks=false,pdfborder={0 0 1},backref=false,colorlinks=true]
 {hyperref}
\hypersetup{
pdfauthor={Clifford Cheung},
 citecolor=black,linkcolor=black,urlcolor=black}

\newcommand{\appendixref}[1]{\hyperref[#1]{appendix~\ref{#1}}}
\def\equationautorefname~#1\null{eq.\,(#1)\null}
\usepackage{breakurl}
\usepackage{breakurl}
\usepackage[hang,flushmargin]{footmisc} 
\allowdisplaybreaks
\makeatletter

\usepackage{etoolbox}
\apptocmd{\thebibliography}{\justifying\setlength{\leftskip}{7.4mm}}{}{} 
 
 \usepackage{relsize}
\usepackage{babel}

\makeatletter
\def\simgt{\mathrel{\lower2.5pt\vbox{\lineskip=0pt\baselineskip=0pt
           \hbox{$>$}\hbox{$\sim$}}}}
\def\simlt{\mathrel{\lower2.5pt\vbox{\lineskip=0pt\baselineskip=0pt
           \hbox{$<$}\hbox{$\sim$}}}}
\makeatother

\usepackage{changepage}

\newcommand{\be}{\begin{equation}}
\newcommand{\ee}{\end{equation}}
\newcommand{\bea}{\begin{eqnarray}}
\newcommand{\eea}{\end{eqnarray}}
\newcommand{\Fig}[1]{Fig.~\ref{#1}}

\newcommand{\Eq}[1]{Eq.~(\ref{#1})}

\newcommand{\App}[1]{App.~\ref{#1}}

\newcommand{\eq}[2]{\be\begin{aligned}#1 \label{#2}\end{aligned}\ee}

\newcolumntype{P}[1]{>{\centering\arraybackslash}p{#1}}

\newcommand{\mL}{m_L}
\newcommand{\mH}{m_H}

\newcommand{\vL}{u_L}
\newcommand{\vH}{u_H}
\newcommand{\zL}{z_L}

\newcommand{\RS}{r_S}
\newcommand{\Rsc}{r_{\Phi}}
\newcommand{\Rvc}{r_{A}}

\newcommand{\yL}{y_L}

\newcommand{\dotbarxL}{\dot{\bar{x}}_L}
\newcommand{\dotbarxH}{\dot{\bar{x}}_H}

\newcommand{\ddotbarxH}{\ddot{\bar{x}}_H}

\newcommand{\barx}{\bar{x}}
\newcommand{\barxL}{\bar{x}_L}
\newcommand{\barxH}{\bar{x}_H}

\newcommand{\dx}{\delta x}

\newcommand{\dg}{\delta g}
\newcommand{\dG}{\delta \Gamma}

\newcommand{\dxL}{\delta x_L}
\newcommand{\dxH}{\delta x_H}

\usepackage[compat=1.1.0]{tikz-feynman}
\usepackage{tikz}
\usetikzlibrary{shapes.geometric}

\tikzset{
bgsource/.style={decorate, thick, empty dot, scale=1.50},
bgraviton/.style={decorate, decoration={complete sines, amplitude=1mm, segment length=2mm, pre length=0mm, post length=0mm}, double, thick},
graviton/.style={decorate, decoration={complete sines, amplitude=1mm, segment length=2mm, pre length=0mm, post length=.2mm}, thick},
recoil/.style={decorate, thick, empty dot, scale=1.25, inner sep = 1pt},
sourceone/.style={decorate, thick, dot, fill=white, scale=1.25, inner sep = 1pt},
binsertion/.style={decorate, thick, crossed dot, scale=1.00},
bscalar/.style={decorate, double, thick}
}

\newcommand{\duffone}{
\begin{tikzpicture}
        \begin{feynman}
          \vertex (f) at (0, 0);
          \vertex[bgsource] (c) at (0,1) {};
          \diagram*{
            (c) -- [graviton] (f) 
          };
        \end{feynman}
    \end{tikzpicture}
}

\newcommand{\dufftwo}{
\begin{tikzpicture}
        \begin{feynman}
          \vertex (f) at (0, 0);
          \vertex (i) at (0,0.5);
          \vertex[bgsource] (c) at (-0.5,1) {};
          \vertex[bgsource] (d) at (0.5,1) {};
          \diagram*{
            (f) -- [graviton] (i) -- [graviton] (c),
            (i) -- [graviton] (d)
          };
        \end{feynman}
    \end{tikzpicture}
}

\newcommand{\duffthreeone}{
\begin{tikzpicture}
        \begin{feynman}
          \vertex (f) at (0, 0);
          \vertex (i) at (0,0.5);
          \vertex[bgsource] (b) at (-0.5,1) {};
          \vertex[bgsource] (c) at (0,1) {};
          \vertex[bgsource] (d) at (0.5,1) {};
          \diagram*{
            (f) -- [graviton] (i) -- [graviton] (b),
            (i) -- [graviton] (c),
            (i) -- [graviton] (d)
          };
        \end{feynman}
    \end{tikzpicture}
}

\newcommand{\duffthreetwo}{
\begin{tikzpicture}
        \begin{feynman}
          \vertex (f) at (0, 0);
          \vertex (i) at (0,0.3);
          \vertex (j) at (0.25,0.6);
          \vertex[bgsource] (b) at (-0.5,1) {};
          \vertex[bgsource] (c) at (0,1) {};
          \vertex[bgsource] (d) at (0.5,1) {};
          \diagram*{
            (f) -- [graviton] (i) -- [graviton] (b),
            (i) -- [graviton] (j) -- [graviton] (c),
            (j) -- [graviton] (d)
          };
        \end{feynman}
    \end{tikzpicture}
}

\newcommand{\bprop}{
\begin{tikzpicture}
        \begin{feynman}
          \vertex[sourceone] (i) at (-1, 0 ) {\tiny{\(L\)}};
          \vertex[sourceone] (f) at (1, 0) {\tiny{\(L\)}}; 
          \diagram*{
            (i) -- [bgraviton, half left]  (f)
          };
        \end{feynman}
    \end{tikzpicture}
    }

\newcommand{\brecoil}{
\begin{tikzpicture}
        \begin{feynman}
          \vertex[recoil] (c) at (0,1.0) {\tiny{\(H\)}};
          \vertex[sourceone] (i) at (-1, 0 ) {\tiny{\(L\)}};
          \vertex[sourceone] (f) at (1, 0) {\tiny{\(L\)}};
          \diagram*{
            (c) -- [bgraviton, quarter right] (i), (c) --  [bgraviton, quarter left] (f),
          };
        \end{feynman}
    \end{tikzpicture}
    }

\newcommand{\bpropscalar}{
\begin{tikzpicture}
        \begin{feynman}
          \vertex[sourceone] (i) at (-1, 0 ) {\tiny{\(L\)}};
          \vertex[sourceone] (f) at (1, 0) {\tiny{\(L\)}}; 
          \diagram*{
            (i) -- [bscalar, half left]  (f)
          };
        \end{feynman}
    \end{tikzpicture}
    }

\newcommand{\lsource}{
\begin{tikzpicture}
        \begin{feynman}
          \vertex (f) at (1, 0);
          \vertex[sourceone] (c) at (0,0) {\tiny{\(L\)}};
          \diagram*{
            (c) -- [bgraviton] (f) 
          };
        \end{feynman}
    \end{tikzpicture}
}

\newcommand{\bgprop}{
\begin{tikzpicture}
        \begin{feynman}
          \vertex (i) at (-1, 0);
          \vertex (f) at (1, 0);
          \diagram*{
            (i) -- [bgraviton]  (f) 
          };
        \end{feynman}
    \end{tikzpicture}
}

\newcommand{\flatprop}{
\begin{tikzpicture}
        \begin{feynman}
          \vertex (i) at (-1, 0);
          \vertex (f) at (1, 0);
          \diagram*{
            (i) -- [graviton]  (f) 
          };
        \end{feynman}
    \end{tikzpicture}
}

\newcommand{\propone}{
\begin{tikzpicture}
        \begin{feynman}
          \vertex (i) at (-1, 0);
          \vertex (f) at (1, 0);
          \vertex[binsertion] (c) at (0,0) {};
          \diagram*{
            (i) -- [graviton](c) -- [graviton]  (f) 
          };
        \end{feynman}
    \end{tikzpicture}
}

\newcommand{\proptwo}{
\begin{tikzpicture}
        \begin{feynman}
          \vertex (i) at (-1, 0);
          \vertex (f) at (1, 0);
          \vertex[binsertion] (c) at (-0.34,0) {};
          \vertex[binsertion] (d) at (0.34,0) {};
          \diagram*{
           (i) -- [graviton] (c) -- [graviton] (d) -- [graviton]  (f) 
          };
        \end{feynman}
    \end{tikzpicture}
}

\newcommand{\brecoilvert}{
\begin{tikzpicture}
        \begin{feynman}
          \vertex (i) at (-1, 0);
          \vertex (f) at (1, 0);
          \vertex[recoil] (c) at (0,0) {\tiny{\(H\)}};
          \diagram*{
            (c) -- [bgraviton] (i),
            (c) -- [bgraviton] (f) 
          };
        \end{feynman}
    \end{tikzpicture}
}

\begin{document}

\title{Effective Field Theory for Extreme Mass Ratios}

\author{Clifford Cheung}
\affiliation{Walter Burke Institute for Theoretical Physics, California Institute of Technology, Pasadena, CA 91125}
\author{Julio Parra-Martinez}
\affiliation{Walter Burke Institute for Theoretical Physics, California Institute of Technology, Pasadena, CA 91125}
\affiliation{Department of Physics and Astronomy, University of British Columbia, Vancouver, V6T 1Z1, Canada}
\author{Ira Z. Rothstein}
\affiliation{Department of Physics, Carnegie Mellon University, Pittsburgh, PA 15213}
\author{Nabha Shah}
\affiliation{Walter Burke Institute for Theoretical Physics, California Institute of Technology, Pasadena, CA 91125}
\author{Jordan Wilson-Gerow}
\affiliation{Walter Burke Institute for Theoretical Physics, California Institute of Technology, Pasadena, CA 91125}

\begin{abstract}
\noindent 

We derive an effective field theory describing a pair of gravitationally interacting point particles in an expansion in their mass ratio, also known as the self-force (SF) expansion.  The 0SF dynamics are trivially obtained to all orders in Newton's constant by the geodesic motion of the light body in a Schwarzschild background encoding the gravitational field of the heavy body.  The corrections at 1SF and higher are generated by perturbations about this configuration---that is, the geodesic deviation of the light body and the fluctuation graviton---but crucially supplemented by an operator describing the recoil of the heavy body as it interacts with the smaller companion.  Using this formalism we compute new results at third post-Minkowskian order for the conservative dynamics of a system of gravitationally interacting massive particles coupled to a set of additional scalar and vector fields.

\end{abstract}

\preprint{CALT-TH 2023-035}

\maketitle

\preprint{}

\maketitle

\noindent {\bf Introduction.} The landmark observation of gravitational waves by LIGO and Virgo \cite{LIGO} has sparked a  scientific revolution in the fields of astrophysics and gravitation.  The recent discovery of a stochastic gravitational wave background by NANOGrav \cite{NANOGrav} has added fuel to this fire.

In parallel with these longstanding experimental efforts is a decades-long theoretical program to derive the \textit{ab initio} predictions of general relativity (GR).  This endeavor has generated numerous lines of attack, including numerical relativity (NR) \cite{Pretorius:2005gq, Lehner_2014,Cardoso_2015}, effective one body (EOB) theory \cite{Buonanno:1998gg}, self-force (SF) methods \cite{PoissonReview,PoundReview,BarackReview}, and perturbative 
post-Newtonian (PN) calculations using traditional methods in GR \cite{Blanchet2014} and effective field theory (EFT) \cite{NRGR}.   More recently, the modern scattering amplitudes program \cite{ElvangHuang, dixon2013-review, cheung2017tasi,  Chetyrkin:1981qh, Henn:2013pwa, Kosower:2018adc} has been retooled towards these efforts in what is known as the post-Minkowskian (PM) expansion \cite{Bertotti1960,Westpfahl1979,Westpfahl1985,Smatrix,Damour2018-pmeob2, Cheung2018-2PM,Bern:2019nnu,Bern:2021dqo,Bern:2021yeh, Herrmann:2021tct, 
DiVecchia:2021bdo,
Herrmann:2021lqe,
Brandhuber:2021eyq, 
Manohar:2022dea, Bjerrum-Bohr:2022blt, Kalin:2020mvi, Dlapa:2021npj, Dlapa:2021vgp, dlapa2023-letter,  Mogull:2020sak, jakobsen2023dissipative, DiVecchia:2023frv}.
Of course, all of these approaches are highly complementary (see e.g., \cite{Barack:2010ny, Nagar:2022fep, Detweiler:2008ft, Barack:2011ed, Bini:2013zaa, 
Damour:2014afa,
vandeMeent:2016hel, 
Antonelli:2019ytb,
Bini:2019nra, 
Bini:2020wpo, Gralla:2021qaf, Long:2021ufh, 
Khalil:2022ylj,
Barack:2022pde, Barack:2023oqp, Whittall:2023xjp, Adamo:2023cfp, Galley_2009}).

Looking to the future, we can expect new insights into the physics of ultra-compact binaries and extreme mass ratio inspirals (EMRIs) from the proposed LISA experiment \cite{LISA1,LISA2}.  However, the EMRI problem is intractable in the vast majority of theoretical approaches, including NR, PN, and PM.   The only extant theoretical approach to the EMRI problem is SF theory, which uses a hybrid of analytical and numerical approaches to calculate the motion of a small companion orbiting a much heavier body, as an expansion in their mass ratio,
\eq{
\lambda = \frac{\mL}{\mH}.
}{}  
In SF theory, the light body induces a perturbation of the black hole spacetime, which then reacts back onto the light body, and so on and so forth (cf.~some reviews \cite{PoissonReview,PoundReview,BarackReview} and state of the art computations at ${\cal O}(\lambda)$, or 1SF, for generic bound orbits in Kerr~\cite{vandeMeent-eccentric,vandeMeent-generic} and at ${\cal O}(\lambda^2)$, or 2SF, for quasicircular orbits in Schwarzschild~\cite{Pound2SF,Warburton2SF,Wardell2SF}).

In this letter we revisit the classic problem of describing the dynamics of a heavy and light body as a systematic expansion in their mass ratio, $\lambda$. Notably, in the limit of vanishing $\lambda$, many physical systems are analytically {\it solvable}.  For example, in the textbook solution to the Rutherford scattering problem, high-energy alpha particles impinge on a gold nucleus described entirely by a rigid $1/r$ background.  Of course, this description is only valid in the limit that the gold nucleus is infinitely heavy---which begs the question, how does one systematically compute the leading nontrivial correction in $\lambda$?  Can this effect be encoded as an {\it operator} added to the theory of a $1/r$ background?  

Analogous logic holds for gravity.  The limit of vanishing mass ratio corresponds to the 0SF dynamics of a probe particle in geodesic motion on a Schwarzschild background.  These background dynamics are understood analytically to all orders in perturbation theory.  But is there an {\it operator} that encodes the 1SF order corrections, and so on and so forth?

Here we offer an affirmative answer to this question in the context of massive, gravitationally interacting {\it point particles}.  At 1SF order, the dynamics are described by a standard background field theory of a light particle worldline coupled to graviton fluctuations in a Schwarzschild background,  supplemented by a ``recoil operator'' encoding the wobble of the background black hole induced by its interactions with the light particle:
\eq{
S_{\rm recoil} &=   -\frac{\mH}{2} \int d\tau \,    \dG^\mu_{H}(\barxH) \frac{1}{\partial_\tau^2}  \dG_{H \mu}(\barxH) \, .
}{recoil_op_GR}
Here we have defined $  \dG^\mu_{H}(\barxH)= \dotbarxH^{\alpha} \dotbarxH^{\beta} \dG^\mu_{\;\;\alpha\beta}(\barxH)$ and $\dG^\mu_{\;\;\alpha\beta}=\Gamma^\mu_{\;\;\alpha\beta}-\bar\Gamma^\mu_{\;\;\alpha\beta}$ is the difference between the connection and its background value.  Physically, \Eq{recoil_op_GR} is generated by integrating out the fluctuations of the heavy body trajectory at 1SF.  In our framework, the 2SF and higher order analogs of the recoil operator can be systematically computed as well.

The key insight in our analysis is to reinterpret the known analytic formulas of classical GR---the Schwarzschild metric and the geodesics of probe particles---as an implicit resummation of an infinite class of Feynman diagrams in {\it flat space} graviton perturbation theory \cite{Duff, Smatrix, DdimSchw}.   In particular, the Schwarzschild background, $\bar g_{\mu\nu}(x) = \eta_{\mu\nu} + \bar\gamma_{\mu\nu}(x)$, is given by the infinite sum of flat-space Feynman diagrams that compute the graviton one-point function induced by a heavy particle:
\eq{
    \bar\gamma_{\mu\nu}
    = \vcenter{\hbox{\duffone}} + \vcenter{\hbox{\dufftwo}} + \vcenter{\hbox{\duffthreeone}} + \vcenter{\hbox{\duffthreetwo}} + \ \ \cdots 
}{Duff}
Similarly, the trajectories of geodesics implicitly encode an infinite set of flat-space Feynman diagrams describing interactions of light and heavy particles.

Of course, it is far easier to manipulate known analytic solutions in classical GR than to build them order by order in perturbation theory.  For this reason, we exactly invert the sequence of logic of \cite{Duff}, and use the Schwarzschild metric and known geodesic trajectories to extract perturbative information.  In doing so, we can perform PM calculations in a  streamlined way that should pay dividends at high PM orders.

Crucially, having ascertained which flat space Feynman diagrams are resummed by the dynamics of a classical probe, we immediately see that there are missing contributions---and at 1SF order these are entirely accounted for by the recoil operator in \Eq{recoil_op_GR}.  Furthermore, by explicitly framing classical GR in terms of flat space perturbation theory, we can use standard tools such as dimensional regularization to deal with point-like or self-energy divergences.

Applying our framework---which at 1SF is simply the background field method plus a recoil operator---we derive old and new results governing conservative gravitational dynamics up to 3PM accuracy.   In particular, we compute the radial action for two massive, gravitationally interacting particles without spin, including the effects of scalar or vector fields which are coupled directly to the light body.   While the results of the present work are limited to 1SF gravity, a longer forthcoming work~\cite{long_paper} will contain many technical details on the systematic derivation of the EFT at 1SF, 2SF, and beyond, as well as applications to non-gravitational theories. 

\medskip

\noindent {\bf Basic Setup.} We begin with the Einstein-Hilbert action coupled to a pair of massive particles \cite{Damour:1995kt, Kalin:2020mvi, Mogull:2020sak},
\eq{
S =& 
-\frac{1}{16\pi G}  \int d^4x  \sqrt{-g} R -  \! \!\sum_{i=L,H}  \frac{m_i}{2}  \int d\tau \, \dot x_i^\mu \dot x_i^\nu g_{\mu\nu}(x_i)  \, ,
}{S_GR}
where we work in mostly minus metric throughout. Here we have gauge fixed the einbein to unity, so $\dot x_i^2=1$.

Our construction is based on expanding the trajectories and metric about their 0SF values,
\eq{
x^\mu_i  = \bar x^\mu_i + \dx_i^\mu \qquad \textrm{and} \qquad
g_{\mu\nu} = \bar g_{\mu\nu} + \dg_{\mu\nu} \, .
}{pert_GR}
Here $\bar x^\mu_i \sim \bar g_{\mu\nu} \sim {\cal O}(\lambda^0)$ are {\it explicit functions} describing 0SF dynamics of a probe in a Schwarzschild background.   Meanwhile, $\dx_i^\mu \sim \dg_{\mu\nu}\sim {\cal O}(\lambda^1)$ are {\it dynamical modes} controlling all contributions at 1SF and higher \footnote{Note that the action itself can have explicit powers of $\lambda$, e.g.~in the contribution to \Eq{S_GR} from the light particle.}.  In particular, $\dx_i^\mu$ is the deviation from geodesic motion, while $\dg_{\mu\nu}$ is the graviton perturbation propagating in a Schwarzschild background.  Inserting \Eq{pert_GR} into \Eq{S_GR}, the action becomes literally that of the background field method, so $S=S_{\rm BF}[\bar g, \delta g,\barxL, \dxL,\barxH, \dxH]$.

An important conceptual point now arises. In standard SF theory, the totality of the dynamics are described by a background field action $S_{\rm BF}[\bar g, \delta g,\barxL, \dxL]$ in which the only degrees of freedom are the geodesic deviation of the light worldline, $\dxL^\mu$, and the fluctuation graviton, $\dg_{\mu\nu}$.   How can such a theory emerge from our starting point of a pair of gravitationally interacting point particles, which we have just shown is described by $S_{\rm BF}[\bar g, \delta g,\barxL, \dxL,\barxH, \dxH]$?  What happened to the heavy particle?
Furthermore, how do we make sense of singular self-force contributions in background field method such as $\bar g_{\mu\nu}(\bar{x}_H)$?  As we will see, while the standard SF theory is perfectly fine, these naive confusions can be explained and handled quite simply in our setup.

\medskip

\noindent {\bf Background Field Theory as Resummation.}  Let us first consider the dynamics at 0SF order, corresponding to an infinite mass ratio between the heavy and light particles.   In this limit the heavy particle is undeflected, so it travels on an inertial trajectory,
\eq{
\barxH^\mu(\tau) = \vH^\mu \tau \, ,
}{}
while serving as a point source for the background gravitational field, which is the boosted Schwarzschild metric.

Famously, this metric can be computed order by order using graviton perturbation theory about flat space \cite{Duff}.  Concretely, the Feynman diagrams shown in \Eq{Duff} form the one-point function of the flat space graviton, which is equal to the difference between the Schwarzschild metric and the flat metric. The precise choice of coordinates for the resulting metric are dictated by the choice of field basis and gauge fixing in the original flat space perturbative formulation.  This essential procedure has been refined and reformulated in many contexts, for instance using modern amplitudes methods \cite{Smatrix, DdimSchw}.
In the present work, we will use the boosted Schwarzschild metric in isotropic gauge,
\eq{
\bar g_{\mu\nu}(x) =  f_+(r)^4 \eta_{\mu\nu} + \left[ \frac{f_-(r)^2}{f_+(r)^2} - f_+(r)^4 \right]\vH{}_\mu \vH{}_\nu \, ,
}{Schwarzschild}
where $f_\pm(r) = 1\pm\tfrac{r_S}{4r}$.  Here $r=\sqrt{(\vH x)^2-x^2}$ is the boosted radial distance from the black hole center  and $r_S  = 2G \mH$ is the Schwarzschild radius.  Expanding the metric in powers of Newton's constant,
 \eq{
\bar\gamma_{\mu\nu}(x)\;\;\;=& \phantom{{}+{}} \frac{r_S}{r}  (\eta_{\mu\nu} -2 \vH{}_\mu  \vH{}_\nu)\\
&+ \frac{1}{8} \left(\frac{r_S}{r}\right)^2  (3 \eta_{\mu\nu} + \vH{}_\mu  \vH{}_\nu) +\cdots
\, ,
}{bkgd_GR}
we obtain the one-point function of the flat space graviton at all orders in the PM expansion.

Note that none of this implies that the Schwarzschild solution of GR is not a {\it vacuum} solution.  Rather, the claim is that the PM expansion of the Schwarzschild metric coincides order by order with the gravitational field of an inertial point source.   

Now let us move on to the light particle, whose 0SF trajectory is a probe geodesic in a Schwarzschild spacetime. While these trajectories are analytically soluble \cite{ChandrasekharBook, Barack2022}, their closed-form expressions are better known in parametric form, i.e.~$r(\theta)$, rather than in explicit time domain, $i.e$ $r(\tau)$ and $\theta(\tau)$.  Nevertheless, starting from the known parametric solutions one can mechanically extract the time domain expression for the probe trajectory order by order in the PM expansion,
$\barxL^\mu = \sum_{k=0}^\infty \barx_k^\mu$,
where the first few terms in isotropic coordinates are
\eq{
\bar{x}_{0}^{\mu}=&\; b^{\mu}+ \vL^{\mu}\tau \\
\bar{x}_{1}^{\mu}=&\; 
\frac{\RS\,  \textrm{arcsinh}\left(\frac{v\tau}{b}\right) \left(\sigma  (2 v^2-1) \vH^{\mu}+\vL^{\mu}\right)}{2v^3}\\
&-\frac{\RS  (2 v^2+1) \sqrt{b^2 + v^{2} \tau^2} \, b^{\mu}}{2b^2 v^2} \, .
}{time_domain_traj1}

Here $\sigma=u_Hu_L$, $v=(\sigma^{2}-1)^{1/2}$, $b=\sqrt{|b_{\mu}b^{\mu}|}$ is the impact parameter, and the trajectories have been computed with time-symmetric boundary conditions. As we will see, the above 0PM and 1PM expressions are sufficient to compute up to 3PM in the conservative dynamics. As described in \App{app:Feynman}, these geodesic trajectories can be used to directly extract data about corresponding perturbative Feynman diagrams in flat space. 

In summary, known background metrics and geodesic trajectories can be straightforwardly distilled into remarkably simple PM integrands for perturbative calculations at higher orders. This will be explained in detail in Ref.~\cite{long_paper}.

\medskip

\noindent {\bf Black Hole Recoil.}  The last puzzle piece is the {\it recoil} of the black hole background.  When we center a static Schwarzschild black hole at the origin, it stays there for all eternity, past and future.  As a result, the background metric is secretly defined in the {\it rest frame} of the physical black hole, which need not be an inertial frame.  Indeed, at 1SF order, the light particle should induce a nonzero deflection of the heavy particle.

In the context of standard SF theory, this subtle effect is accounted for by the non-radiating part of the metric perturbations, captured in the low multipole moments \cite{Zerilli1970, Detweiler2004} whose computation requires imposing suitable boundary conditions at the event horizon.

Since our starting point is a point particle effective field theory, there is an easier way: simply compute the path integral over the geodesic deviation of the heavy particle, $\dxH^\mu$.  At 1SF order this is a Gaussian integral which can be done exactly, yielding exactly the recoil operator in \Eq{recoil_op_GR}.  The corrections at 2SF order and higher are similarly computed in a mechanical fashion.

To summarize, the background field theory defined by a light particle worldline interacting with fluctuation gravitons in a Schwarzschild background is {\it not} equivalent to a theory of heavy and light particles interacting gravitationally.  To match the latter, one must supplement the former with the recoil operator.  Doing so yields our main result, which is the action for our EFT of extreme mass ratios at 1SF order,
\eq{
S_{\rm EFT} &= S_{\rm BF}[\bar g, \delta g,\barxL, \dxL] + S_{\rm recoil}.
}{SEFT}
This quantity is precisely the standard background field action for the light particle interacting with fluctuating gravitons in a Schwarzschild background, plus the recoil operator in  \Eq{recoil_op_GR}.  See \App{app:Feynman} for a summary of the Feynman rules in this EFT.  Note that a key advantage of the recoil operator is that it obviates the need to iteratively solve for the deflection of the heavy particle and its concomitant corrections to the background spacetime.

Since the recoil operator has a $1/\partial_\tau^2$ pole, it requires a choice of boundary conditions,  which corresponds to an $i\epsilon$ prescription for the $\dxH^\mu$ propagator.
For the present work, we only consider conservative scattering, for which the propagator for the heavy particle fluctuation never goes on-shell and the choice of $i\epsilon$ is thus immaterial (see \cite{long_paper} for details).   Here we use the Feynman prescription for simplicity, but more generally one should deploy the recoil operator with the $i \epsilon$ prescription appropriate to the physical process in question.

From the viewpoint of traditional EFT, it is somewhat peculiar to integrate out states to generate a {\it nonlocal} operator.  One usually restricts consideration to a kinematic regime in which such effects becomes effectively local---and indeed for the conservative region this is the case.  Still, one might reasonably worry about power counting ambiguities stemming from these nonlocal interactions.  Nevertheless, since the masses of particles are effectively coupling constants in the worldline formalism, higher order insertions of nonlocal operators such as the recoil operator are suppressed in the SF expansion.

\medskip
\noindent {\bf Self-Energy and Regularization.} Our derivation of the recoil operator in \Eq{recoil_op_GR} actually entails a critical subtlety involving self-energy and its regularization.  Strictly speaking, the equation for the heavy particle geodesic in its own Schwarzschild background is
\eq{
\ddotbarxH^\mu  +  \bar\Gamma^{\mu}_{\;\; \alpha\beta}(\barxH) \dotbarxH^\alpha \dotbarxH^\beta =0 \, .
}{}
The second term is singular because it involves the background connection at the position of the heavy particle.

In traditional SF theory, self-energy divergences of this type afflict the light particle dynamics.  However, in the present context of point particle effective field theory, all such  divergences are trivially eliminated using dimensional regularization.  In particular, any self-energy divergence ultimately arises in perturbation theory as a self-energy diagram. For example, the self-energy contribution from the Newton potential arises from the $r\rightarrow 0$ limit of $1/r \sim\int d^{3-2\epsilon} q \, e^{iqr}/q^2 $.  However, taking the limit inside the integral yields a scaleless integrand that vanishes \emph{by definition} in dimensional regularization.

The upshot here is that we can simply drop all PM corrections involving the background metric evaluated at $\barxH$, so dimensional regularization effectively sets $\bar g_{\mu\nu}(\barxH) = \eta_{\mu\nu} $ and $ \bar\Gamma^{\mu}_{\;\; \alpha\beta}(\barxH)  = \bar R_{\mu\nu\alpha\beta}(\barxH)  = 0$.
While these equations may appear at odds with general covariance, they are not.  Rather, the statement is that formally $\bar g_{\mu\nu}(\barxH)$ is generally covariant, but its difference from $\eta_{\mu\nu}$ at the point $\barxH$ is zero in dimensional regularization.  Note that numerous terms have been dropped in the recoil operator in \Eq{recoil_op_GR} on account of dimensional regularization.
Furthermore, the indices of the recoil operator are implicitly contracted with flat space metrics, rather than the divergent background metric.  As we will see later on, this prescription exactly yields known correct results for the conservative dynamics.

\medskip

\medskip
\noindent {\bf Gravitational Scattering with Additional Fields.} Our effective field theory can also be used to compute new results at 1SF order.  As it turns out, the SF theory community has a particular interest in a model in which  the gravitational action in \Eq{S_GR} is supplemented with a massless scalar or vector which couple directly to the light particle but interact only gravitationally with the heavy particle  \cite[eg.][]{Burko2000,Barack2022}. This is a well-known toy model for the full gravitational SF problem. Thus we add to \Eq{S_GR} the additional scalar and vector contributions,
\begin{align}\label{scalar-vect}
S_{\Phi, A}=&
\int d^4 x  \sqrt{- g}  \left[ \tfrac12 \nabla_\mu \Phi \nabla^\mu \Phi + \tfrac{1}{2} \xi R \Phi^2  -\tfrac14 F_{\mu\nu} F^{\mu\nu} \right] \nonumber \\
&- \mL \! \int \!d\tau\, \left[  \yL \Phi(x_L)+\zL  A_{\mu}(x_L) \dot x_L^\mu\right]\, . 
\end{align}
For generality we have included a non-minimal coupling of the scalar to gravity.

Since the heavy particle does not couple directly to the new fields,  the recoil operator is unaffected. However,  new contributions to scattering are induced by the gravitational interactions of the scalar and vector fields sourced by the light particle.  At 1SF these arise solely from the background field diagrams.

\begin{figure}
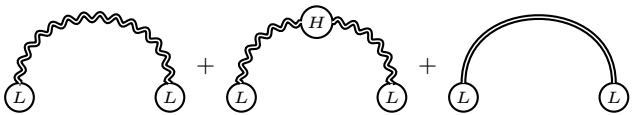

    $\hbox{\bprop} \mathrel{\raisebox{4ex}{$\!+\!$}} \hbox{\brecoil} \mathrel{\raisebox{4ex}{$\!+\!$}} \hbox{\bpropscalar}$
    \caption{Background-field Feynman diagrams contributing to the 1SF radial action. The circles denote the light geodesic source ($L$) and the heavy recoil operator ($H$). The double lines denote background field propagators for the graviton (wavy) and the scalar or vector fields (straight). }
    \label{fig:3PM}
\end{figure}

\medskip
\noindent {\bf Calculation of the Radial Action.}  As a consistency check of our formalism, we will calculate the radial action for conservative dynamics 
at 1SF order.
The radial action is a convenient gauge-invariant quantity encoding the conservative scattering dynamics.  Moreover, it is a generating function for the time-delay and scattering angle. 
Common approaches to the PM radial action involve applying simple maps to scattering amplitudes~\cite{Bern:2021dqo, Bern:2021yeh} or to momentum impulses~\cite{Kalin:2019rwq, Kalin:2019inp}. Here, we will instead directly calculate the radial action by evaluating the ``on-shell'' value of the EFT action in \Eq{SEFT}, which physically corresponds to plugging in the solutions to the classical equation of motion.   The resulting object is the radial action, which takes the form
\eq{
S_{\rm EFT}\big|_{\textrm{on-shell}}=\RS\mL\sum_{n=0}\lambda^{n}I_n,
}{S_onshell}
where $n$ labels the order in the SF expansion and each $I_n$ contains all orders in $G$. In what follows we perturbatively expand the 1SF contribution $I_1$ in powers of $G$.

In the language of quantum field theory, \Eq{S_onshell} is calculated by evaluating the path integral over all the particle and graviton degrees of freedom.  At 1SF order, the radial action is equal to the sum of connected tree diagrams in which the light body sources a graviton, which propagates in the full Schwarzschild background, has an arbitrary number of recoil operator insertions, and then returns to the light body.  These manipulations are equivalent to the Feynman rules in \App{app:Feynman}.  

Notably, the resulting tree diagrams effectively generate loop integrals arising from Fourier transforms in the worldline trajectories.
These integrals can be easily evaluated in dimensional regularization using integration-by-parts (IBP) identities \cite{Chetyrkin:1981qh} and canonical differential equations \cite{Henn:2013pwa}, as explained e.g. in \cite{Parra-Martinez:2020dzs}.
Since we are focusing on conservative dynamics, and up to 3PM order there are no tail effects to handle, we can expand all loop momenta in the potential region. Diagrams with more than one recoil operator insertion vanish in the potential region, and thus the 1SF dynamics up to 3PM order are computed by  the two diagrams in \Fig{fig:3PM}. At higher PM orders tail effects mandate that we supplement these diagrams by the appropriate number of recoil operator insertions while extending the loop integration to include contributions from the ultrasoft region.

The final result for the 1SF-3PM radial action, including scalar and vector contributions, is
\begin{widetext}

\eq{
I_1 =&\phantom{{}+{}} \frac{\RS}{b}\frac{3 \pi}{16} \frac{ 5 \sigma ^2-1}{\sqrt{\sigma ^2-1}}
+ \frac{\RS^2}{b^2} \left(\frac{ \sigma  \left(36 \sigma ^6-114 \sigma ^4+132 \sigma ^2-55\right)}{12\left(\sigma
   ^2-1\right)^{5/2}}
   +\frac{\left(-4 \sigma ^4+12 \sigma ^2+3\right) \textrm{arccosh}\,\sigma }{2\left(\sigma ^2-1\right)}\right) 
   \\
   &
   -\frac{\Rvc}{ b} \frac{\pi}{8}\frac{ 3 \sigma ^2-1}{\sqrt{\sigma ^2-1}}  -\frac{\RS\Rvc}{b^{2}} \left(\frac{\sigma  \left(8 \sigma ^4-28 \sigma ^2+23\right)}{12 \left(\sigma
   ^2-1\right)^{3/2}}
   +\frac{ \left(2 \sigma ^2+1\right) \textrm{arccosh}\,\sigma }{\left(\sigma ^2-1\right)}\right) 
   \\
   &
   - \frac{\Rsc}{b} \frac{\pi}{8}\frac{\sigma ^2-1+4 \xi}{ \sqrt{\sigma ^2-1}}  -\frac{\RS\Rsc}{b^{2}}\frac{ \sigma  \left(2 \sigma ^4-\sigma ^2-1+\xi  \left(6 \sigma ^2-3\right)\right)}{6 \left(\sigma
   ^2-1\right)^{3/2}}
}{}
\end{widetext}
where we've introduced the scalar and vector charge radii, $\Rsc=\yL^2 \mH/(4\pi)$, $
\Rvc=\zL^2 \mH/(4\pi) $.

This is in agreement with the radial action inferred from scattering angles previously reported in the literature \cite{Bern:2019nnu, Gralla:2021qaf}. The 3PM scalar result with $\xi = 0$ agrees with the recently reported result derived via scattering amplitudes~\cite{Barack:2023oqp}, while the 3PM vector result is new. The 2PM results pass an additional check, which is that taking $\lambda=1$ gives the 2PM radial action in the probe limit for trajectories in certain charged black hole backgrounds~\cite{long_paper}.

A highly nontrivial check of our EFT is that the 2SF-3PM radial action must agree with the 0SF-3PM radial action \cite{Damour:2019lcq}. Although we have not presented the 2SF Feynman rules, they are straightforwardly derived by expanding the action one further order in the mass ratio. To check consistency, we have indeed carried out this expansion and computed the complete 2SF-3PM radial action for the model above, including gravitation as well as the toy scalar and photon. The result exactly matches the probe limit in the appropriate background as required for consistency of our approach~\cite{long_paper}.

\medskip
\noindent {\bf Conclusions.} Our EFT framework leaves numerous avenues for future exploration.  First and foremost, it is of utmost importance to see if there is a direct relation, if any, to current approaches to black hole recoil in standard SF theory, which involve a so-called ``matter-mediated'' force \cite{Pfenning:2000zf}.  Perhaps by connecting our results with those ideas, our EFT can have a more direct application to existing SF results.

Secondly, it would be interesting to generalize our results beyond the case of Schwarzschild, which corresponds to a heavy, spinless, minimally coupled particle.  Here a natural extension would be to include spin by considering a Kerr background.  Another option would be to study a background sourced by a neutron star, or an electrically charged particle.  In all of these cases, the recoil operator will change, simply because the propagation of the heavy particle will be modified.

Third, it should be relatively straightforward to apply our EFT to the nonconservative sector, i.e., to the dynamics of gravitational radiation.  Since the recoil operator is simply a nonlocal in time correction to the graviton propagator, it can be readily included in any PM calculation for graviton emission.

Last but not least, we should note that the general approach of this work---that we can systematically derive corrections to background field method from flat space perturbation theory---can also be applied outside the context of gravity.  Here a natural target is the the study of fluid mechanics.  In this case the long-range force carrier is the fluid velocity, the backgrounds are classical solutions to the Navier-Stokes equations, and the probe particles are worldlines that are minimally coupled to the fluid.

\vspace{3mm}

\noindent {\it Acknowledgments:} While this paper was at a late stage we learned of the upcoming work~\cite{Kosmopoulos:2023bwc} on a framework for self-force using scattering amplitudes in curved space. We thank the authors for coordinating release of their work. C.C., N.S., and J.W.-G. are supported by the Department of Energy (Grant No.~DE-SC0011632) and by the Walter Burke Institute for Theoretical Physics. J.W.-G is also supported by a Presidential Postdoctoral Fellowship and the Simons Foundation (Award Number
568762). I.Z.R. is supported by the Department of Energy (Grant No.~DE-FG02-04ER41338 and FG02-06ER41449).

\appendix

\section{Feynman rules}

\label{app:Feynman}

In this section we summarize the Feynman rules for the EFT defined in \Eq{SEFT}, which describes a light particle worldline coupled to fluctuating gravitons in a Schwarzschild background---with the addition of the recoil operator.   Interpreted as a background-field action, \Eq{SEFT} has a corresponding set of background-field Feynman diagrams that can be used to compute the 1SF radial action, e.g.~as depicted in \Fig{fig:3PM}.  To compute in the PM expansion, however, it is natural to further expand these background-field Feynman diagrams order by order in Newton's constant.  In this picture the fundamental perturbation theory is in flat space, and the difference of the Schwarzschild metric and particle geodesics from flat space and straight lines, respectively, are considered PM corrections.

To begin, let us define the background-field effective action governing the light particle worldline and the fluctuation graviton in a curved background,
\begin{align} 
\label{SBFexplicit}
&S_{\rm BF}[\bar g, \dg,\barxL, \dxL] + S_{\rm GF}= S[\bar g, ,\barxL] \\
& + \int_x   \sqrt{-\bar g} \left[ \dg_{\mu\nu} \bar T_L^{\mu\nu} +
\tfrac{1}{16\pi G}\left(-\tfrac14  \dg_{\mu\nu}\bar\nabla^2 \dg^{\mu\nu}  + \tfrac18   \dg \bar\nabla^2 \dg\right. \right.  \nonumber \\
&   -\tfrac12 \dg_{\mu\nu} \dg_{\rho\sigma} \bar R^{\mu\rho\nu\sigma} -
 \tfrac12 (\dg_{\mu\rho} \dg_\nu^{\;\;\rho}  - \dg_{\mu\nu} \dg )\bar R^{\mu\nu}   \nonumber \\
& +\left.\left. \! \tfrac14 (\dg_{\mu\nu} \dg^{\mu\nu} -\tfrac12 \dg^2)\bar R \right)\right] + \cdots\,, \nonumber
\end{align}
where we have added a Lorenz gauge fixing term $S_{\rm GF} = \tfrac{1}{32\pi G} \int_x \sqrt{-\bar g} F_\mu F^\mu $ with $F_\mu = \bar\nabla^\nu \delta g_{\mu\nu} -\tfrac12 \bar\nabla_\mu \delta g$. In the first line of \Eq{SBFexplicit}, the quantity $S[\bar g, ,\barxL]$ is just the probe radial action, which as usual is computed by plugging in the background metric and light geodesic in Eq.~\eqref{S_GR}.  Starting at the second line of \Eq{SBFexplicit}, we show the terms needed to compute 1SF corrections, where the ellipses denote the higher order corrections. 

Here $\bar T_L^{\mu\nu}$ is the stress-energy tensor for the geodesic trajectory of the light particle, corresponding to a source term which implies the momentum-space Feynman rule,
\begin{align}
& \vcenter{\hbox{\lsource}}  \!=\! \sqrt{-\bar g}\bar T_L^{\mu\nu}(p) \!= \! \lambda  \mH \!  \int \!d\tau \, e^{-ip \barxL}  \dotbarxL^\mu\dotbarxL^\nu \\
&=\lambda  \mH e^{-ipb} \left( \vL^\mu  \vL^\nu \delta(\vL p) + 2i {\cal O}_{\alpha}^{\;\;\, \mu \nu}(\vL,p)\barx_{1}^\alpha(\vL p)\right) \nonumber \\
&+\cdots  \, , \nonumber
\end{align}
where the ellipsis denotes higher PM orders and where we have defined
\eq{
{\cal O}^{\alpha \mu\nu}(u_{j},p) \! =\!\frac12 ( (u_{j}^\mu \eta^{\nu\alpha} + u_{j}^\nu \eta^{\mu \alpha}) (u_{j} p) -  u_{j}^\mu u_{j}^\nu p^\alpha )\,.
}{} 
together with the frequency-domain trajectory, $ \barx^\mu_i(\omega)=\int d\tau \, e^{-i\omega\tau} \barx_i^\mu(\tau) $, and we have also explicitly expanded to subleading order.  Case in point, we can trivially recast the trajectories in \Eq{time_domain_traj1} into the form of perturbative Feynman diagrams.  Concretely, using identities such as
\eq{
\textrm{arcsinh}\left(\frac{v\tau}{b}\right) &=  \frac{1}{\partial_\tau}\left( \frac{v }{(b^2+ v^2 \tau^2)^{1/2}} \right) \, ,
}{arcsinh_simp}
we can write the 1PM time-domain trajectory as
 \eq{
 \barx_1^\mu &= -\frac{\RS}{2v^2 b^2}(2v^2+1)b^{\mu}(b^{2}+v^{2}\tau^{2})^{1/2}\\
 &+\frac{\RS}{2v^{2}}\left(\sigma(2v^{2}-1)\vH^{\mu} +\vL^{\mu}\right)\frac{1}{\partial_\tau}\frac{1}{(b^{2}+v^{2}\tau^{2})^{1/2}} \, .
}{}
Powers of the spatial distance $(b^2+v^2\tau^2)^{1/2}$ can be rewritten as simple fourier integrals. In the frequency domain, the trajectory then has a simple form
\eq{
\barx_1^\mu(\omega)&=\frac{i(2\pi)^3\RS}{v^{2}}\int\frac{d^{4}k}{(2\pi)^{4}}e^{-ik\bar{x}_0}\delta(\omega-\vL k)\delta(\vH k)\\
&\times\left(\frac{8(2v^{2}+1)\Pi^{\mu\nu}}{b^{2}}\frac{k_{\nu}}{(k^{2})^{3}}-\frac{(\sigma(2v^{2}-1)\vH^{\mu}+\vL^{\mu})}{k^{2}(\vL k)}\right),
}{}
where $\Pi^{\mu\nu}=\eta^{\mu\nu}-v^{-2}(\sigma \vH^{\mu}-\vL^{\mu})\vL^{\nu}-v^{-2}(\sigma \vL^{\mu}-\vH^{\mu})\vH^{\nu}$ projects onto directions orthogonal to the four-velocities. Note that factor of $\omega^{-2}$ which one would expect from perturbatively solving the geodesic equation has been reduced in power to $\omega^{-1}$ in one term, and entirely eliminated in the other. 

The full background-field propagator can be expanded perturbatively around flat space as
\begin{align}
    \vcenter{\hbox{\bgprop}}
    & = \vcenter{\hbox{\flatprop}} +
    \vcenter{\hbox{\propone}} \nonumber \\
    & + \vcenter{\hbox{\proptwo}} + \ \ \cdots
\end{align}
where the circles denote background field insertions and the leading term is
 the flat-space graviton propagator,
\begin{equation}\label{deDonder}
\!\!\flatprop  \;\;= \;\; \frac{16 \pi G i}{p^2} \left( \eta_{\mu\rho}\eta_{\nu\sigma} \!+\!\eta_{\mu\sigma}\eta_{\nu\rho}\! - \!\eta_{\mu\nu} \eta_{\rho\sigma}\right) \, ,
\end{equation}
which is in de Donder gauge on account of our choosing Lorenz gauge the original background-field action defined in \Eq{SBFexplicit}.

Of course, the true background is the Schwarzschild metric, but in the PM expansion we can treat these effects as order by order corrections to the flat space graviton two-point function. These contributions are obtained by taking the difference of the isotropic gauge Schwarzschild metric from the flat space metric and expanding in PM, which is simply the momentum-space version of \Eq{bkgd_GR},
\begin{align} 
\bar \gamma_{\mu\nu}(p)  = &  - \frac{8\pi G \mH }{p^2} ( \eta_{\mu\nu}-2\vH{}_\mu \vH{}_\nu ) \delta(\vH p) \phantom{\Bigg|}\\ 
& -  \frac{8\pi^2 G^2 \mH^2}{\sqrt{-p^2}}  (3\eta_{\mu\nu}+\vH{}_\mu \vH{}_\nu ) \delta(\vH p)
+\cdots   \, .  \phantom{\Bigg|}\nonumber
\end{align}
The corresponding insertion is just the three-point vertex,  from standard graviton perturbation theory in flat space, connecting two graviton lines to a linearized background metric. Note the appearance of non-zero curvatures in Eq.~\eqref{SBFexplicit}, which also appear as insertions. These arise because the metric is not a vacuum solution but is sourced by the heavy particle. 

At low PM orders, we only need the background and geodesics to linear order and hence the background field method is not more efficient than performing a flat space perturbative calculation. However, at higher orders one sees considerable simplification, since in isotropic gauge the background metric insertions are simple powers in the radius $r$ whose Fourier transforms yield very simple dependencies on the momentum transfer $q$ induced by the insertion.  In particular, the resulting Feynman rules are the same as for simple loop integrands with numerator structures that depend solely on $\eta_{\mu\nu}$ and $\vH{}_{\mu}\vH{}_{\nu}$ and are thus effectively scalar. Hence, the background field method effectively performs tensor reduction on subdiagrams within multiloop Feynman diagrams.

Finally, as explained in the main text, the background field action must be supplemented by the recoil operator in \Eq{recoil_op_GR}. It is trivial to compute the corresponding two-point vertex, which is
\begin{align} \label{recoil_vertex_GR}
&\vcenter{\hbox{\brecoilvert}} \\
 &= {  \frac{i\mH}{2}   \frac{\delta(\vH p_1+ \vH p_2)}{(\vH p_1)(\vH p_2)} {\cal O}^{\alpha \mu_1 \nu_1}(\vH, p_1)  {\cal O}_\alpha^{\;\; \mu_2 \nu_2}(\vH, p_2) } \,  .\nonumber
\end{align}
The above Feynman rules are sufficient to compute the 1SF radial action for point-like compact bodies  order by order in the PM expansion. 

\vspace{-5mm}

\bibliographystyle{utphys-modified}

\bibliography{shortEFT}

\end{document}